\documentclass[nofootinbib, superscriptaddress, notitlepage, twocolumn, floatfix]{revtex4-2}
\usepackage{graphicx,color}
\usepackage[section]{placeins}
\usepackage{textgreek,stackengine}
\usepackage{amsmath,amssymb,bm}
\usepackage[left=2cm, right=2cm, top=3cm]{geometry}
\graphicspath{ {Images/} }
\usepackage{amsthm}
\usepackage{textcomp}
\usepackage{mathtools}
\usepackage{algorithm}
\usepackage[noend]{algpseudocode}
\usepackage{dsfont}
\usepackage{braket}
\usepackage{appendix}
\usepackage{xcolor}
\definecolor{mediumcarmine}{rgb}{0.69, 0.25, 0.21}
\usepackage[colorlinks=true, citecolor=blue, linkcolor=mediumcarmine]{hyperref}
\usepackage{xpatch}
\usepackage{enumitem}
\usepackage{tikz}
\usetikzlibrary{arrows}
\usepackage{xcolor}
\usepackage[caption=false]{subfig}
\usepackage{graphicx}

\usepackage[utf8]{inputenc}
\usepackage[T1]{fontenc}
\usepackage{csquotes}
\usepackage[english]{babel}
\usepackage{amsmath}
\usepackage{amsthm}
\usepackage{amssymb}
\usepackage{mathtools}
\usepackage{empheq}
\usepackage{dsfont}
\usepackage{bm}
\usepackage{xspace}
\usepackage{algorithm}

\theoremstyle{plain}

\theoremstyle{definition}

\theoremstyle{plain}

\theoremstyle{plain}

\makeatletter
\xpatchcmd{\algorithmic}{\itemsep\z@}{\itemsep=2ex plus2pt}{}{}
\makeatother

\begin{document}

\title{Randomized Benchmarking with Stabilizer Verification and Gate Synthesis}

\author{E.~Derbyshire}
\affiliation{School of Informatics, University of Edinburgh, Edinburgh EH8 9AB, UK}

\author{R.~Mezher} 
\affiliation{School of Informatics, University of Edinburgh, Edinburgh EH8 9AB, UK}

\author{T.~Kapourniotis}
\affiliation{School of Informatics, University of Edinburgh, Edinburgh EH8 9AB, UK}
\affiliation{Department of Physics, University of Warwick, Coventry CV4 7AL, UK}

\author{E.~Kashefi}
\affiliation{School of Informatics, University of Edinburgh, Edinburgh EH8 9AB, UK}
\affiliation{Laboratoire d\textquotesingle Informatique de Paris 6, CNRS, Sorbonne Universit\'{e}, 4
place Jussieu, 75005 Paris, France}

\begin{abstract}
 Recently, there has been an emergence of useful applications for noisy intermediate-scale quantum (NISQ) devices notably, though not exclusively, in the fields of quantum machine learning and variational quantum algorithms. In such applications,  circuits of various depths and composed of different sets of gates are run on NISQ devices. Therefore, it is crucial to find practical ways to capture the general performance of circuits on these devices. Motivated by this pressing need, we modified the standard Clifford randomized benchmarking (RB) and interleaved RB schemes targeting them to hardware limitations. Firstly we remove the requirement for, and assumptions on, the inverse operator, in Clifford RB by incorporating a tehchnique from quantum verification. This  introduces another figure of merit by which to assess the quality of the NISQ hardware, namely the acceptance probability of quantum verification. Many quantum algorithms, that provide an advantage over classical algorithms, demand the use of Clifford as well as non-Clifford gates. Therefore, as our second contribution we develop a technique for characterising a variety of non-Clifford gates, by combining tools from gate synthesis with interleaved RB. Both of our techniques are most relevant when used in conjunction with RB schemes that benchmark generators (or native gates) of the Clifford group, and in low error regimes.
\end{abstract}
\maketitle

\section{Introduction}
Of fundamental importance for near-term intermediate scale quantum (NISQ) devices is being able to efficiently characterise noise when performing a number of computations of varying complexity and depth. Comprehensive techniques of noise characterisation aid in understanding the dominant sources of noise plaguing different quantum hardware platforms \cite{superconducting,iontraps,photonics} and consequently contribute to developing methods of noise mitigation \cite{EBL+18}. Noise characterisation methods such as \cite{WE+16, FW+20, MBZ+21} aim to reconstruct the noise channel of quantum systems for specific cases such as making assumptions about the locality of noise \cite{MBZ+21}. Techniques such as Quantum Tomography \cite{MGS+13, BKF+20, CPF+10, AMP+03} and Direct Fidelity Estimation (DFE) \cite{FTL+11} aim to find the degree of noise in a particular state/gate-set/process with the resources for tomography scaling exponentially with the system size. Whilst the methods of Refs.~\cite{WE+16, FW+20, MBZ+21} and DFE are more efficient than most tomography schemes, they can still be very costly in practice. In most cases, these methods do not account for errors from state preparation and measurement (SPAM). Quantum verification is another branch of device characterisation that usually determines whether a quantum circuit/state is acceptably close to the desired one based on how well it passes certain tests set by a so-called verifier \cite{GKK19}. It comes with notions of security that the previous characterisation techniques do not and while it scales polynomially with the system size, in practice it requires a significant overhead of physical qubits, circuit depth, interaction with classical computers, quantum communication and assumptions about the capabilities of the verifier. In practice, one does not always require the complete information about the most general noise channel, especially if the noise of a NISQ device behaves somewhat uniformly across different gates and experimental runs. On the contrary, it is useful to determine more general quantities, such as the average gate error, for more realistic types of noise. This is the key motivation behind another approach to characterising noise which  we  expand on in this work,  the experimental  technique  of  Randomized Benchmarking (RB) \cite{KLR+08, MGE+11, MGE+12}. 

RB involves implementing many random sequences of quantum gates, and it provides a relatively efficient way to determine the average performance of an $n$-qubit gate-set $\mathcal{G} \subset \mathcal{U}(2^n)$ (the $n$-qubit unitary group) on a specific quantum hardware, where the gates are sampled according to a probability distribution $\mu$. A key difference between RB and the previously mentioned partial noise characterisation techniques is that it is robust to errors from state preparation and measurement (SPAM), making it highly desirable in practice. Most RB protocols require the couple $\{\mu, \mathcal{G}\}$  to form an exact unitary 2-design \cite{DCE+09} in order to reduce the noise channel, acting on the gates, to a simpler form; although there are alternatives to this rule such as character RB \cite{HXV+19}. The output is a figure of merit $r$ which is a measure of the average error of the gate-set; the noise channel is often assumed gate and time-independent, however these restrictions have been relaxed in some cases \cite{MGE+12}. Standard RB methods require the implementation of the inverse of a random sequence of gates in a short time-step, and make the assumption that the noise channel on this inverse gate is on average the same for all gates and all sequences. In this work, we focus on Clifford RB, where gates are drawn from the uniform distribution over the $n$-qubit Clifford group. Clifford RB, although scalable, is not especially informative beyond a few qubits. This is because the $n$-qubit Clifford elements that are characterised are compiled from the native gates of a quantum hardware. In \cite{FH+18, PCR+19} the authors modified the RB technique to allow for noise characterisation of these native gates. Unfortunately, these modifications still rely on some impractical assumptions, see Sec.~\ref{sec: RB} for further details and an overview of standard RB and App.~\ref{app: RB} for a technical summary.

In our first contribution we remove the inverse step and its assumptions, by adapting a specific technique in quantum verification \cite{GKK19}, the Stabilizer Verification (SV) protocol of Markham and Krause \cite{MK+20}, and performing local Pauli measurements on multiple repetitions of a single sequence; we call our method Randomized Benchmarking with Stabilizer Verification (RBSV). By removing the inverse step, our work removes the unrealistic assumptions in \cite{FH+18,PCR+19}, whilst still being compatible with standard RB protocols; moreover, RBSV is very easily combined with both of these generator RB methods \cite{FH+18, PCR+19}. Through RBSV, a parameter $r_{rbsv}$ can be extracted which is close to the average error $r$ in high fidelity regimes. RBSV also introduces a new figure of merit to RB, with the acceptance probability from the SV protocol \cite{MK+20}, which lower bounds the sequence fidelity and acts as a failure signature such that if it is not high enough, then the experimentalist should not run the RBSV protocol until the noise is lowered. Furthermore, our technique can be generalised, due to the nature of verification techniques \cite{GKK19}, to scenarios where the noise is gate and time-dependent. These added features make the RBSV protocol potentially more useful for benchmarking NISQ hardware than other techniques which also manage to remove the inverse step in RB such as Monte Carlo interleaved RB \cite{CRK+17}.

The motivation was to perform Clifford RB without the inverse step, making only \emph{small} modifications such that the method would still be simple to perform and straightforward to interpret. We perform numerical simulations for simple models of depolarising noise, and observe that our techniques give more accurate results as compared with Clifford RB in regimes where the noise strengths are low, which we would expect due to constraints from verification \cite{GKK19}. Arguably, this makes our technique more tailored to near term  quantum devices where one usually runs circuits composed of high fidelity gates on a low number of (high quality) qubits \cite{P+18}. We first introduce the RBSV protocol in Sec.~\ref{sec: RBSV} presenting numerical simulations for a two-qubit system comparing results from RB and RBSV in Sec.~\ref{sec: simulations}. We then find the conditions on the measurement errors required for our protocol to deliver accurate results (Sec.~\ref{sec: imperfectmeas}), and finally discuss the computational resources required to accurately perform RBSV in Sec.~\ref{sec: scaling}.
 
Generalising RB to non-Clifford gates is a widely researched area \cite{HXV+19, DYM+20, CWE+15, BE+18, CMB+16, GCM+12, HF+17} and we expand on some of the ideas presented in these works. Most of these techniques require that the gate-set benchmarked is a finite group \cite{HXV+19, CWE+15}, a sub-set of the Clifford group \cite{BE+18, CMB+16, GCM+12} or in the case of \cite{DYM+20} that the gate-set is an $\epsilon$-approximate 2-design. In our second contribution, we use techniques from gate synthesis \cite{MDM+13} as a way to characterise specific non-Clifford gates that can complement the Clifford group to achieve universal quantum computation. In particular, we present a special case of interleaved RB \cite{MGJ+12} where the fixed Clifford element is synthesised such that it contains Clifford gates and non-Clifford gates; the latter placed in non-trivially cancelling positions. We require the number of non-Clifford gates in each fixed element to be low. We call our technique Interleaved Randomized Benchmarking with Gate Synthesis (iRB+GS) in Sec.~\ref{sec: iRB+GS}. This is a simple modification of standard interleaved RB, however it opens up the potential to characterise a broad variety of non-Clifford gates and to our knowledge, it has not yet been explored in other theoretical works. We first present the form of iRB+GS for a synthesis involving two non-Clifford gates and in Sec.~\ref{sec: noiseanalysis} we analyse the error on the estimate of the average error per non-Clifford in this case. In Sec.~\ref{sec: construction} we present a specific synthesis construction for the self-inverting generating set of the two-qubit Clifford group, where the non-Clifford is the controlled-Phase gate. Finally, in Sec.~\ref{sec: rotations} we generalise our proofs to a parametric family of controlled-Phase(k) gates, for rotations $2\pi/2^k$, where the synthesis involves more than two non-Clifford gates.

\section{Background 1.} 
\subsection{Randomized Benchmarking}\label{sec: RB}
There are many variants of RB \cite{HRO+20} but most RB methods generally involve running $K_m$ random sequences, of various sequence lengths $m$, of gates randomly sampled from a gate-set $\{\mathcal{U}\}$, and an additional gate $m+1$ also from $\{\mathcal{U}\}$ that inverts the preceding sequence. Where $K_m$ and $m$ are positive integers. The underlying theory of RB dictates that the probability distribution over the gate-set $\{\mathcal{U}\}$ is (normally) an exact 2-design \cite{DCE+09}, which allows for any noise channel acting on the gates to be reduced to a depolarising channel; a channel that is simple and easy to characterise. The noise on each gate composes as $\Lambda_{U_i} = \Lambda_i \circ U_i, U_i \in \{\mathcal{U}\}$, and the inverse gate is assumed to act in a single step with a single error channel that is the same on average for each sequence. The probability of the initial state surviving the process, referred to as the survival probability, which if perfectly implemented would be the Identity, is measured  for each random sequence $j \in \{1, ..., K_m\}$ of sequence length $m$. The average survival probability $P_m$ for each sequence length is found by averaging over all $K_m$ random sequences at length $m$. The data $P_m$ is plot against $m$ and fit to a decay curve, which for gate and time-independent noise is as follows:
\begin{equation}
    P_m = A_0 + B_0p^m \enspace ,
\end{equation}
where the fit parameters $A_0$ and $B_0$ absorb errors from state preparation and measurement and $p$ is the depolarising parameter, of the equivalent depolarising channel, that is extracted and used to calculate the RB parameter $r = (d-1)(1-p)/d$ where $d = 2^n$ is the dimension of the Hilbert space. In the presence of gate and time-independent noise we model $\Lambda_{U_i} = \Lambda$ and $r = 1 - F(\Lambda_{U}, U)$ the average gate-set infidelity where $F(\Lambda_{U}, U) = F(\Lambda, I) = \int_{\rho} Tr(\rho \Lambda(\rho))$ is the average fidelity over all pure states of an average gate $U \in \mathcal{U}$. For further technical details, please see App.~\ref{app: RB}.
\par 
One of the most important examples of an exact 2-design, and the one that we focus on in this paper, is the uniform distribution over the $n$-qubit Clifford group $\mathcal{C}_n$ \cite{DCE+09}. Clifford RB is still one of the simplest methods of the RB field, and is straightforward to interpret. One of the drawbacks of the Clifford RB method is that it provides a figure of merit $r$ \emph{only} for the errors affecting a \emph{group element} of $\mathcal{C}_n$. The group elements are in practice compiled from a set of \emph{native} gates which are the building blocks of the quantum hardware being tested, so there is not necessarily a simple interpretation of the parameter $r$ with respect to the underlying gates. The methods of Refs.~\cite{FH+18, PCR+19} address this issues, introducing ways to benchmark native gates of a quantum hardware. However, these adaptations rely on somewhat unrealistic assumptions such as access to a general Clifford measurement \cite{PCR+19}, or the ability to implement an arbitrary group element of $\mathcal{C}_n$ in a single time step \cite{FH+18}. The single inverse step is one of the key components of Clifford (and other) RB methods. The assumptions on the inverse are not unrealistic for very small-scale systems but when benchmarking larger systems, non-Clifford gates or generators of the Clifford group they become far less feasible. In the seminal paper on RB \cite{KLR+08} Knill et al performed a Pauli rotation after a sequence of random gates on one qubit, that ensures the final state is in an eigenstate of $Z$ and suggested using rapid mixing of generators for multi-qubit RB rather than the full Clifford group. Subsequent extensions to \cite{KLR+08} sought to make the method more scalable and efficient, which is why an inverse step was introduced. More recently, Chassaur et al \cite{CRK+17} introduced Monte Carlo sampling to approximate the fidelity of the sequences using the techniques of \cite{FL+11, SLP+11}, rather than implementing the inverse step. We remove the inverse step by applying the technique of stabilizer verification and our method remains close in form to standard Clifford RB.

\subsection{Stabilizer Verification}\label{sec: SV}
This method is presented as graph state verification in \cite{MK+20} however the property of a graph state being uniquely defined by a set of stabilizers that is required for this protocol is shared by any stabilizer state, and therefore the proofs and results in \cite{MK+20} hold for any stabilizer state, which we apply it to. Therefore, we refer to it as stabilizer verification.

Any (pure) $n$-qubit stabilizer state $\ket{\psi}$ can be written as a Clifford circuit acting on the zero state: $\ket{\psi} := C_{m}, \cdots , C_{1} \ket{0}^{\otimes n}$. Where $C_i$  are elements of the $n$-qubit Clifford group, $C_i\in \mathcal{C}_n$.  A \emph{stabilizer} $s$ of an $n$-qubit stabilizer state $\ket{\psi}$ is a unitary operator which leaves $\ket{\psi}$ invariant: $s\ket{\psi} = \ket{\psi}$, that is $s$ \emph{stabilizes} $\ket{\psi}$. 

The stabilizer $s$ is always a tensor product of $n$-single qubit Pauli's from the set $\{I,X,Y,Z\}$ with some phase $\pm 1$ or $\pm i$. In other words, $s$ is a member of the $n$-qubit Pauli group $\mathcal{P}_n$. The set of all stabilizers of $|\psi\rangle$ forms a group, denoted as $\mathcal{S}:=\{s \in \mathcal{P}_n | s|\psi\rangle=|\psi\rangle\}$. This group is called the \emph{stabilizer group} of $|\psi\rangle$.

Markham et al \cite{MK+20} present a verification method for certifying stabilizer states which we summarise here and call Stabilzer Verification (SV). Suppose we are given as input $R$ copies of an $n$-qubit quantum state which, in the ideal case, should be $R$ copies of  $\rho=|\psi\rangle \langle \psi |$. We denote the overall system of $(R.n)$ qubits as $\sigma$, and in the ideal case $\sigma=\rho^{\otimes R} = (|\psi\rangle\langle\psi|)^{\otimes R}$. The protocol consists first of choosing at random $(R-1)$ copies, and measuring for each copy $i = 1,...,R-1$ a randomly chosen element $s_i \in \mathcal{S}$. Therefore, performing local Pauli measurements (or the Identity) on each of the $n$ qubits of each copy, such that the overall effect amounts to measuring the stabilizer $s_i$. The possible measurement outcomes of a local $X,Y$ or $Z$ Pauli measurement are $\{+1,-1\}$ and the measurement outcome of an $s_i$ measurement is simply the product of measurement outcomes of all (non-identity) Pauli's constituting $s_i$. The stabilizer measurement $s_i$ is \emph{successful} when the outcome of an $s_i$ measurement is $+1$. Note that, for an ideal copy $\rho=|\psi\rangle\langle \psi|$, the measurement of any $s_i \in \mathcal{S}$ will always be successful, since $|\psi\rangle$ is the simultaneous $+1$ eigenstate of all elements of $\mathcal{S}$.The SV protocol \emph{accepts} if all the stabilizer measurements on all $(R-1)$ copies are successful.

Let $\rho_{out}$ denote the quantum output of the protocol, i.e. the state of the remaining unmeasured copy tensored with an auxiliary qubit which encodes the end result of the protocol, and let $\{|\text{\text{Acc}}\rangle, |\text{Rej}\rangle\}$ be orthogonal quantum states encoding the end result of the protocol, i.e. whether it accepts or rejects. The output state $\rho_{out}$ can be written as $$\rho_{out}=P_{\text{acc}}\rho_{acc} \otimes |\text{\text{Acc}}\rangle \langle \text{\text{Acc}}|+(1-P_{\text{acc}})\rho_{\text{rej}} \otimes |\text{Rej}\rangle \langle \text{Rej}|  \enspace .$$
 Where $P_{\text{acc}}$ is the acceptance probability of the protocol, and $\rho_{acc}$ and $\rho_{rej}$ are the states of the unmeasured copy conditioned on accepting and rejecting respectively, averaged over all possible choices of $M-1$ copies and stabilizers.
 The goal of this verification, and indeed any verification, is to ensure that $\rho_{acc}$ is as close as possible to the ideal state $\rho=|\psi\rangle \langle \psi|$.
 In \cite{MK+20}, it is shown that for any input state $\sigma$: 
 \begin{equation}
 Tr((1-|\psi\rangle \langle \psi| \otimes |\text{Acc}\rangle \langle \text{Acc}|)\rho_{out}) \leq \dfrac{1}{R} \enspace .
 \end{equation}
 When we substitute the expression of $\rho_{out}$ into the above, we have:
 \begin{center}
 \begin{equation}
 \begin{split}
     &Tr(P_{\text{acc}}((1-|\psi\rangle \langle \psi|)\rho_{acc} \otimes |\text{Acc}\rangle \langle \text{Acc}|)\\
     &=Tr(P_{\text{acc}}((1-|\psi\rangle \langle \psi|)\rho_{acc}))\cdot Tr(|\text{Acc}\rangle \langle \text{Acc}|)
     \\&=P_{\text{acc}}Tr((1-|\psi\rangle \langle \psi|)\rho_{acc}) \leq \dfrac{1}{R} \enspace .
\end{split}
 \end{equation}
 \end{center}
 As mentioned previously, and defined in \cite{MK+20}, $\rho_{acc}$ is the output state, in the case where the protocol accepts, \emph{averaged} over all possible choices of stabilizers and copies. Let $\rho_{acc_k}$ be the output of the protocol conditioned on acceptance for a particular choice $k$ of $R-1$ copies $i = 1, ..., M$ and $M-1$ stabilizers $s_i$. We can write $\rho_{acc}$ as:
 \begin{equation}
     \rho_{acc} = \frac{1}{R \cdot |\mathcal{S}|^{R-1}} \sum\limits_{k = 1, ..., R \cdot |\mathcal{S}|^{R-1}} \rho_{acc_k} \enspace . \nonumber 
\end{equation}
 By the linearity of the trace, and the fact that $Tr((1-|\psi\rangle \langle \psi|) \rho_{acc_k})\geq 0$, we have that for any $k$,
 $$P_{\text{acc}}Tr((1-|\psi\rangle \langle \psi|)\rho_{acc_k}) \leq \dfrac{1}{R} \enspace .$$ 
 Noticing that the trace $Tr(|\psi\rangle \langle \psi|\rho_{acc_k})$ is equal to the Fidelity $F(|\psi\rangle \langle \psi|,\rho_{acc_k})$ and again, using the linearity of the trace, we find that: 
 \begin{equation}
 \label{eq: fidelitysv}
     F(|\psi\rangle \langle \psi|,\rho_{acc_k}) \geq 1-\dfrac{1}{P_{\text{acc}} \cdot R} \enspace .
 \end{equation}
We use the above expression to develop our RB protocol, where we estimate the acceptance probability to find the lower bound on the average fidelity. In our protocol, we measure \emph{all} R copies, accepting if all $R$ stabilizer measurements are successful. This does not change the protocol, as the fidelities we are interested in are those for any randomly chosen copy. The SV protocol provides a natural robustness to preparation errors, because copies need not be identical amongst themselves. 

\section{Randomized Benchmarking with Stabilizer Verification}\label{sec: RBSV}
Extending RB, we introduce the RBSV protocol~\ref{alg: RBSV} for benchmarking the Clifford group using stabilizer verification. As a first step, we assume perfect measurements and that our quantum device is initiated in the state $\rho_\psi$ where, if perfectly implemented, $\rho_\psi = (\ket{0}\bra{0})^{\otimes n}$. In Sec.~\ref{sec: imperfectmeas} we analyse the protocol in the presence of imperfect measurements. All Clifford elements applied are chosen uniformly at random from $\mathcal{C}_n$, where the sequence length $m$ is the number of Clifford elements applied \footnote{Here, when we say Clifford \emph{elements} we are referring to an $n$ qubit Clifford element of the Clifford group, which can contain many multi-qubit gates. For a sequence length $m$, we therefore mean a circuit containing $m$ Clifford elements.}. We represent noisy Clifford elements as $\tilde{C}_{j_i}= \Lambda_{j_i} \circ C_{j_i}$ where $j_i$ indicates the $j_{th}$ random sequence and $i$ it's position in the sequence, i.e. $i = 1, ..., m$, and therefore we define a random sequence of noisy Cliffords as $\tilde{C}_{j,m}$. 

For the remainder of this paper, we will make the assumption that all noise, including SPAM, is independent and identically distributed (i.i.d) between all of the copies $R$ of a specific random sequence $\tilde{C}_{j,m}$, and between each repetition $N_m$ of this same sequence. We assume that the noise is gate and time-independent; this can be extended (as other RB protocols have been \cite{MGE+12}) though not for complex noise models. We model the noise by fixed quantum channels for gate $\Lambda_{i_j} = \Lambda$, preparation $\Lambda_p$ and measurement $\Lambda_m$ which represent the average noise. Note that our protocol could be  generalised to relax this i.i.d assumption, as relaxing this assumption is an inherent property of  verification protocols, however we do not attempt this here.
\begin{figure}[!ht]
\begin{algorithm}[H]
\floatname{algorithm}{Protocol}
\caption{Randomized Benchmarking with Stabilizer Verification}
\label{alg: RBSV}
\renewcommand{\thealgorithm}{}
\begin{algorithmic}[1]
 \State \label{rbsvstep1} For a chosen sequence length $m$, randomly sample $K_m$ sequences labelled by $j \in \{1, K_m\}$ consisting of randomly chosen Clifford gates $C_{j_m} \in \mathcal{C}_n$.
 \State \label{rbsvstep2}For each sequence $C_{j,m} = {C}_{j_m} \cdots {C}_{j_1}$, determine with a classical computer the stabilizer group $S_{j,m}$ of $C_{j,m} \ket{0}^{\otimes n}$ (using the Gottesman-Knill theorem).
 \State \label{rbsvstep3} Implement the sequence $\tilde{C}_{j,m} = \tilde{C}_{j_m} \cdots \tilde{C}_{j_1}$ on initial (imperfect) state $\rho_\psi$. Choose a random stabilizer $s$ from $S_{j,m}$, perform and record a stabilizer measurement of $s$ on the output of the system.
\State \label{rbsvstep4} Repeat step 3) $N_m$ times with the same noisy sequence $\tilde{C}_{j,m}$. If for each repetition, the recorded stabilizer measurements were successful, i.e. the product of the outputs of measurements were $+1$, \emph{accept}; otherwise, \emph{reject}. Record the number of times the sequence is \emph{accept}, $N_{acc}$. The \emph{acceptance probability estimate} is:
\begin{equation}
    \tilde{P}_{\text{acc}_{j,m}} = \frac{N_{acc}}{N_m} \nonumber 
\end{equation}
\State \label{rbsvstep5} Assuming i.i.d. noise between copies, estimate the acceptance probability for a stabilizer verification with \emph{R} copies of the sequence $\tilde{C}_{j,m}$ is:
\begin{equation}
    \tilde{P}_{\text{acc}_{j,m}}(R) = \tilde{P}_{\text{acc}_{j,m}}^R
\end{equation}
from here, estimate the quantity: $1 - \frac{1}{R. P}_{\text{acc}_{j,m}(R)}$ inputting our estimate $\tilde{P}_{\text{acc}_{j,m}}(R)$ which lower bounds the average sequence fidelity (see Eq.~\ref{eq: fidelitysv}): 
\begin{equation}
\label{eq: lowerbound}
    \overline{F}_{j,m} \geq (1 - \frac{1}{R. \tilde{P}_{\text{acc}_{j,m}}(R)}) 
\end{equation}
\State \label{rbsvstep7} For each $m$ we estimate the lower bound on the average sequence fidelity, by averaging over all $K_m$ random sequences:
\begin{equation}
    \overline{F}_{m} = \frac{1}{K_m}\sum_{j = 1, \cdots, K_m}\overline{F}_{j,m} \geq \frac{1}{K_m} \sum_{j = 1, \cdots, K_m} (1 - \frac{1}{R. \tilde{P}_{\text{acc}_{j,m}}(R)}) \nonumber
\end{equation}
\State \label{rbsvstep8} Plot the average sequence fidelities $\overline{F}_{m}$ against sequence length $m$ and fit to a predetermined decay curve $F^{m}$. If the noise is gate and time-independent, the curve will be:
\begin{equation}
    F^{m}_{(0)} = A_0 + B_0 p_{rbsv}^m \nonumber 
\end{equation}
where the fitting parameters $A_0$ and $B_0$ absorb the SPAM errors, and we can extract $p_{rbsv}$ to find $r_{rbsv} = (d - 1)(1 - p_{rbsv})/d$ an indication of the average performance of the gates.
\end{algorithmic}
\end{algorithm}
\end{figure}
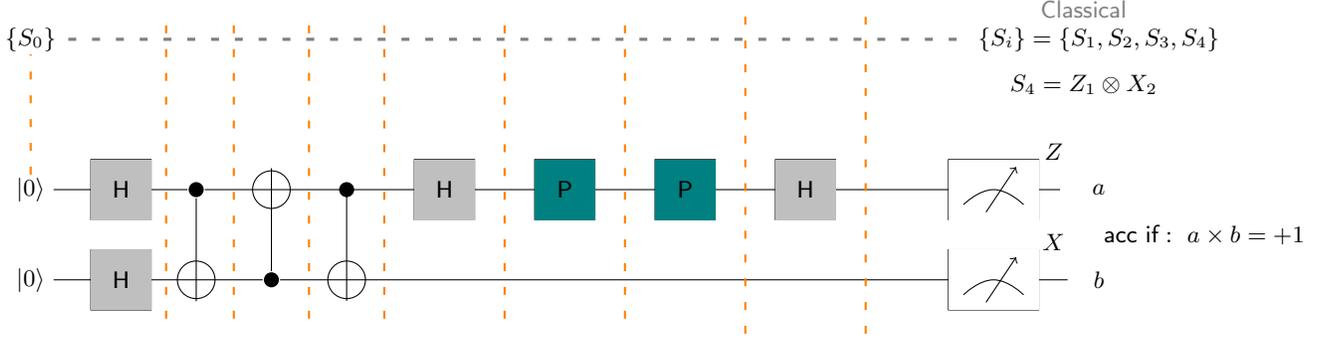
\begin{figure*}[!ht]
    \begin{center}
    \begin{tikzpicture}[font=\sffamily]
\node (v1) at (-6,5) {$\ket{0}$};
\node (v2) at (-5,5) {};
\draw  (v1) edge (v2);
\node (v3) at (-4.6,5) {};
\node (v4) at (-3.8,5) {};
\draw  (v3) edge (v4);
\draw  (-5.2,5.4) rectangle (-4.4,4.6);
\fill [lightgray] (-5.2, 5.4) rectangle (-4.4, 4.6);
\fill[black] (-3.8,5) node (v5) {} ellipse (0.1 and 0.1);
\node  (v10) at (-6,3.8) {$\ket{0}$};
\node (v11) at (-2.8,5) {};
\node (v12) at (-2.8,3.8) {};
\draw (-2.8,5) node(v11) {} ellipse (0.25 and 0.25);
\fill[black]  (v12) ellipse (0.1 and 0.1);
\node (v13) at (-2.8,5.4) {};
\draw (v12) edge (v13);
\node (v14) at (7.8,5) {};
\draw (v5) edge (v14);
\node (v15) at (-3.8,3.8) {};
\fill[black] (-1.8, 5) ellipse (0.1 and 0.1);
\draw (-1.8, 3.8) ellipse (0.25 and 0.25);
\draw (-1.8, 5) edge (-1.8, 3.52);
\draw (-3.8, 3.8) node(v15) {} ellipse (0.25 and 0.25);
\draw  (v10) edge (v12);
\node (v16) at (-3.8,3.4) {};
\draw (v5) edge (v16);
\node (v17) at (-4.8,5) {H};
\node (v6) at (7.9,3.8) {};
\draw  (v12) edge (v6);
\draw  (-5.2,4.2) rectangle (-4.4,3.4);
\fill [lightgray] (-5.2,4.2) rectangle (-4.4,3.4);
\node at (-4.8,3.8) {H};
\draw  (-0.9,5.4) rectangle (-0.1,4.6);
\fill [lightgray] (-0.9,5.4) rectangle (-0.1,4.6);
\node (v20) at (-0.5,5) {H};
\draw (0.7, 5.4) rectangle (1.5, 4.6);
\fill[teal] (0.7, 5.4) rectangle (1.5, 4.6);
\node at (1.1, 5){P};
\draw  (2.3,5.4) rectangle (3.1,4.6);
\fill [teal] (2.3,5.4) rectangle (3.1,4.6);
\node (v21) at (2.7,5) {P};
\draw (3.9, 5.4) rectangle (4.7, 4.6);
\fill[lightgray] (3.9, 5.4) rectangle (4.7, 4.6);
\node at (4.3,5) {H};
\draw  (6.2,5.4) rectangle (7.4,4.6) node (v7) {};
\draw  (v7) rectangle (v7);
\draw  (6.2,4.2) rectangle (7.4,3.4);
\fill [white] (6.2,5.4) rectangle (7.4,4.6);
\fill [white] (6.2,4.2) rectangle (7.4,3.4);
\draw  plot[smooth, tension = .9] coordinates { (7.2,4.8) (6.8,5) (6.4,4.8)};
\draw plot[smooth, tension = .9] coordinates { (7.2,3.6) (6.8,3.8)(6.4,3.6)};
\draw [->] (6.7,4.7)--(7.1,5.3);
\draw [->] (6.7,3.5)--(7.1,4.1);
\node (vj) at (-4.2,7.3) {};
\node (vk) at (-4.2,3) {};
\draw [orange] [thick] [loosely dashed] (vj)--(vk);
\node (vl) at (-3.3,3) {};
\node (vm) at (-3.3,7.3) {};
\draw [orange] [thick] [loosely dashed] (vm)--(vl);
\node (vn) at (-2.3,3) {};
\node (vo) at (-2.3,7.3) {};
\draw [orange] [thick] [loosely dashed] (vo)--(vn);
\node (vp) at (-1.3,3) {};
\node (vq) at (-1.3,7.3) {};
\draw [orange] [thick] [loosely dashed] (vq)--(vp);
\node (vr) at (0.3,3) {};
\node (vs) at (0.3,7.3) {};
\draw [orange] [thick] [loosely dashed] (vs)--(vr);
\node (vt) at (1.9,3) {};
\node (vu) at (1.9,7.3) {};
\draw [orange] [thick] [loosely dashed] (vu)--(vt);
\draw [orange][thick][loosely dashed] (3.5, 7.3)--(3.5, 3);
\draw [orange][thick][loosely dashed] (5.1, 7.3)--(5.1, 3);
\node at (-6,7) {$\{S_0\}$};
\node at (8.2,7) {$\{S_i\} = \{S_1, S_2, S_3, S_4\}$};
\draw [gray] [very thick] [loosely dashed] (-5.5,7)--(6.4,7);
\node at (8,6.4) {$S_4 = Z_1 \otimes X_2$};
\node at (-6,5.2) {};
\node at (-6,6.8) {};
\draw [orange] [thick] [loosely dashed] (-6, 5.2)--(-6, 6.8);
\node at (8.2,5) {$a$};
\node at (8.2,3.8) {$b$};
\node at (9.6,4.4) {$\text{acc} \ \text{if}: \ a \times b = +1$};
\node at (7.6,5.5) {$Z$};
\node at (7.6,4.3) {$X$};
\node at (8,7.4) {\textcolor{gray}{Classical}};
\end{tikzpicture}
    \end{center}
    \caption{An example of the stabilizer measurement scheme for one Clifford element on two qubits. We begin in the state $\ket{00}$ which has stabilizer group $\{S_0\} = \{II, IZ, ZI, ZZ\}$ and as each gate is applied from $G = \braket{H, P, CNOT}$ the stabilizer group gets updated (indicated by the orange dashed line). For this circuit, the final stabilizer group is $\{S_i\} = \{II, ZI, IX, ZX\}$ and we have indicated that if $S_4 = ZX$ was chosen randomly from $\{S_i\}$ then we would measure in the $Z$ basis on qubit 1 and the $X$ basis on qubit 2; accepting only if the product of the outcomes of each measurement is $+1$.}
    \label{fig: rbsvmeasure}
\end{figure*}
\par 
For ease of understanding, the random stabilizer measurement of step \ref{rbsvstep3} of the RBSV protocol~\ref{alg: RBSV} is illustrated in Fig.~\ref{fig: rbsvmeasure}, for one Clifford element only, on a two qubit device. As discussed in Sec.~\ref{sec: SV} the stabilizer group is updated with the update rules from the Gottesman-Knill theorem, where for each stabilizer $s_i$ the updated stabilizer $s_{i+1}$ is equal to $U s_{i} U^{\dagger}$ when a circuit $U$ has been applied to initial stabilizer state $\ket{\psi_i}$. A common way to compute the stabilizer group for larger system sizes is using the tableau representation \cite{AG+04}, however for our simulation results we limit ourselves to two-qubit systems and therefore brute-force the stabilizer updates. 
\par 
The output of the RBSV protocol is the parameter $r_{rbsv}$ which should not be far from the parameter predicted from standard RB which we denote $r_{rb}$, in the high fidelity regime. In the case that the fit parameters are equivalent $r_{rbsv} \geq r_{rb}$, therefore providing an upper bound on the average error per Clifford element. In order for $r_{rbsv}$ to be as close as possible to $r_{rb}$, we need the lower bound on the average Fidelity of sequence $\tilde{C}_{j,m}$: $1 - 1/(R \cdot P_{\text{acc}_{j,m}}(R))$, Eq.~\ref{eq: lowerbound} to be tight, where this expression for the lower bound is directly obtained for our scenario from stabilizer verification \cite{MK+20} (Sec.~\ref{sec: SV}). For RBSV we first estimate the acceptance probability for a single copy (step~\ref{rbsvstep4}) by repeating the same noisy random sequence and measuring $N_m$ times\footnote{If we do not assume i.i.d. noise between copies we would have to repeat the same noisy random sequence $N_m \cdot R$ times}. We do this to reduce the computational overhead for estimating $P_{\text{acc}_{j,m}}(R)$. Since we assume noise is i.i.d. between copies, we have that: $P_{\text{acc}_{j,m}}^{R} = P_{\text{acc}_{j,m}}(R)$.

We want Eq.~\ref{eq: lowerbound} to be tight enough, such that our estimated $\tilde{P}_{\text{acc}_{j,m}}(R)$ can be used to estimate the lower bound on the average sequence fidelity $\overline{F}_{m}$ and subsequently fit to the exponential RB decay curve. The tightness of the lower bound, Eq.~\ref{eq: lowerbound} is not analysed in the stabilizer verification work \cite{MK+20} as it is provided as a way to verify that one has an adequately high enough quality state and not to characterise noise.  We analyse the tightness of the bound in order to determine for what value of $R$ our estimate $\tilde{P}_{\text{acc}_{j,m}}(R) = \tilde{P}_{\text{acc}_{j,m}}^R$ provides the closest value to $\overline{F}_{j,m}$. 
\par 
First, we want to minimise the Drift between the two values, which we define as: 
\begin{equation}
    \text{D}(R) = |(1 - \frac{1}{P_{\text{acc}_{j,m}}^R\cdot R}) - \overline{F}_{j,m}|
\end{equation}
We first determine the rate of change of $\text{D}(R)$ as a function of $R$:
\begin{equation}
    \begin{split}
       \frac{d\text{D}(R)}{dR} &= \frac{P_{\text{acc}_{j,m}}^{R} (R ln(P_{\text{acc}_{j,m}}) +1)}{(P_{\text{acc}_{j,m}}^R\cdot R)^2}\\
    \end{split}
\end{equation}
This rate of change demonstrates that $\text{D}(R)$ increases in the interval of $R \in \{0, 1/ln(\frac{1}{P_{\text{acc}_{j,m}}})\}$ and decreases in the interval of $R \in \{1/ln(\frac{1}{P_{\text{acc}_{j,m}}}), +\infty\}$. Therefore, the closest that $\overline{F}_{j,m}$ and $(1 - \frac{1}{P_{\text{acc}_{j,m}}^R \cdot R})$ get to each other is when:
\begin{equation}
    R = 1/ln(\frac{1}{P_{\text{acc}_{j,m}}})
\end{equation}
The average sequence fidelity $\overline{F}_m$ can be fit to the RB decay curve,
\begin{equation}
    A_0 + B_0 p^m \enspace ,
\end{equation}
and we estimate the average sequence fidelity by using the lower bound:
\begin{equation}
    \overline{F}_m \geq \frac{1}{K_m} \sum\limits_{j = 1, ..., K_m} 1 - \frac{1}{P_{\text{acc}_{j,m}}^R \cdot R} 
\end{equation}
Since we now have the value $R = 1/ln(\frac{1}{P_{\text{acc}_{j,m}}})$ for which the right of the above sum is closest to the average fidelity, we can look at the lower bound of the average sequence fidelity $\overline{F}_m$ from our estimated acceptance probability:
\begin{equation}\label{eq: aveseqfid}
     \overline{F}_m \geq \frac{1}{K_m} \sum\limits_{j = 1, ..., K_m} 1 - e ln(\frac{1}{P_{\text{acc}_{j,m}}}) \enspace , 
\end{equation}
where we have used the fact that $P_{\text{acc}_{j,m}}^{(1/ln(\frac{1}{P_{\text{acc}_{j,m}}}))} = 1/e$, and $e$ is Euler's number, $e \approx 2.71828$.
\par 
In RB, the estimated average sequence fidelity (estimated from the survival probability) is fit to the RB curve: $A_0 + B_0p^m$ in the presence of gate and time-independent noise. In RBSV, we fit to the same curve and we therefore deduce a relationship, based on some assumptions, between the parameters $r_{rb}$ and $r_{rbsv}$ estimated from each protocol by comparing these fits. Let us first assume that for both RB and RBSV we have the same noise model, with the same average strength and SPAM errors. The RB and RBSV estimated average sequence fidelities are fit as follows: 
\begin{equation}
\begin{split}
    P_m &\approx A_{rb} + B_{rb}p_{rb}^m \\
    \frac{1}{K_m}\sum\limits_{j = 1, ..., K_m} 1 -eln(\frac{1}{P_{\text{acc}_{j,m}}}) &\approx A_{rbsv} + B_{rbsv}p_{rbsv}^m \enspace ,
\end{split}  
\end{equation}
where $P_m$ is the experimental survival probability from standard RB and now, we differentiate the parameters with subscripts of $rb$ and $rbsv$ respectively. 

As we know from Eq.~\ref{eq: aveseqfid} the RBSV estimate lower bounds the average sequence fidelity, and we can therefore state that the RBSV estimate also lower bounds the survival probability estimated from RB:
\begin{equation}
\begin{split}
    P_m &\geq \frac{1}{K_m}\sum\limits_{j = 1, ..., K_m} 1 -eln(\frac{1}{P_{\text{acc}_{j,m}}})\\
    A_{rb} +B_{rb}p_{rb}^m &\geq A_{rbsv} + B_{rbsv}p_{rbsv}^m \\
    \frac{A_{rb} - A_{rbsv}}{B_{rb}} + p_{rb}^m &\geq \frac{B_{rbsv}}{B_{rb}} p_{rbsv}^m 
\end{split}
\end{equation}
We make the assumption that, in the high-fidelity regime, $A_{rbsv} \approx A_{rb}$ and $B_{rbsv} \approx B_{rb}$ \footnote{When there are no SPAM errors, we expect this assumption to be more realistic}, and we have:
\begin{equation}
\begin{split}
    p_{rb}^m &\geq p_{rbsv}^m \\
    p_{rb} &\geq p_{rbsv} \\
    1 - p_{rbsv} &\geq 1 - p_{rb} \\
    r_{rbsv} &\geq r_{rb} \\
\end{split}
\end{equation}
Therefore, the parameter $r_{rbsv}$ directly upper bounds the actual average gate-set infidelity $r$:
\begin{equation}
    r_{rbsv} \geq r
\end{equation}
\par

Here we make clear the requirements we have of $P_{\text{acc}_{j,m}}$. For our first bound of Eq.~\ref{eq: lowerbound} to be tight, such that all others follow, we require that $P_{\text{acc}_{j,m}}$ be close to $1$. Therefore, our protocol works best for devices that require high accuracy during computation. Extending RBSV to a regime with a higher volume of, and more complex, noise would be desirable; however, the security and soundness that comes with the verification scheme dictates this limit on $P_{\text{acc}_{j,m}}$ \cite{MK+20}. Though there is marginal flexibility in the value of $P_{\text{acc}_{j,m}}$, the RBSV protocol as it is could not effectively characterise a system with a lot of complex noise sources. In this type of scenario, the estimated value of the acceptance probability $\tilde{P}_{\text{acc}_{j,m}}$ would provide a failure signature, i.e. the amount of noise in the system is too high for the protocol to proceed. 
\subsection{Numerics} \label{sec: simulations}
In order to support our theoretical findings we ran numerical simulations for sequences of lengths $m$ randomly chosen from $m \in \{5, \cdots, 50\}$, with $K_m = 200$ random sequences at each sequence length $m$ and for parameters $R = 1/ln(\frac{1}{\tilde{P}_{\text{acc}_{j,m}}})$ and $N_m = 100$. We ran the standard RB protocol, and the RBSV protocol to compare the output parameters $r_{rbsv}$ and $r_{rb}$. We chose a simple depolarising gate noise channel of the following form:
\begin{equation}
    \Lambda_{\text{dep}} (\rho) = (1-\varepsilon_i)\rho + \varepsilon_i \frac{\mathcal{I}}{d}
\end{equation}
where $d = 2^n$ is the dimension of the Hilbert space, and $\varepsilon_i$ is the average noise strength. For simplicity, we did not model SPAM errors.
\begin{figure}[!ht]
    \begin{center}
    \subfloat[RB and RBSV fit for an average depolarising noise of strength $\varepsilon_i = 0.0001$ (top two lines) and $\varepsilon_i = 0.001$ (bottom two lines)\label{fig: rb_rbsv_0001_001}]{%
    \includegraphics[width=0.45\textwidth]{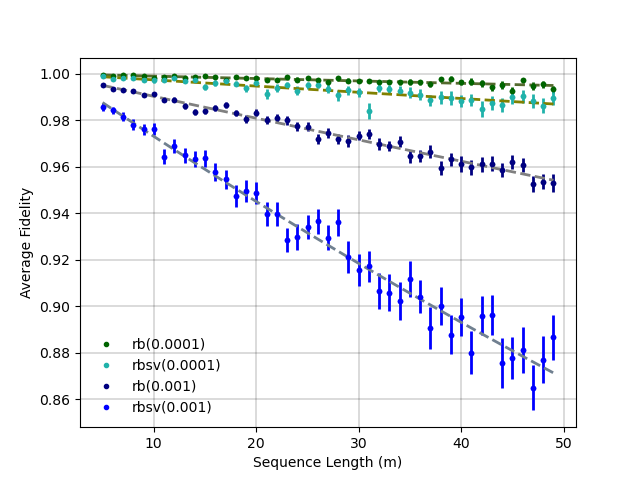}
    }
    \hfill 
    \subfloat[RB and RBSV fidelity comparison for an average depolarising noise of strength $\varepsilon_i = 0.001$ (top two lines) and $\varepsilon_i = 0.005$ (bottom two lines)\label{fig: rb_rbsv_001_005}]{%
    \includegraphics[width = 0.45\textwidth]{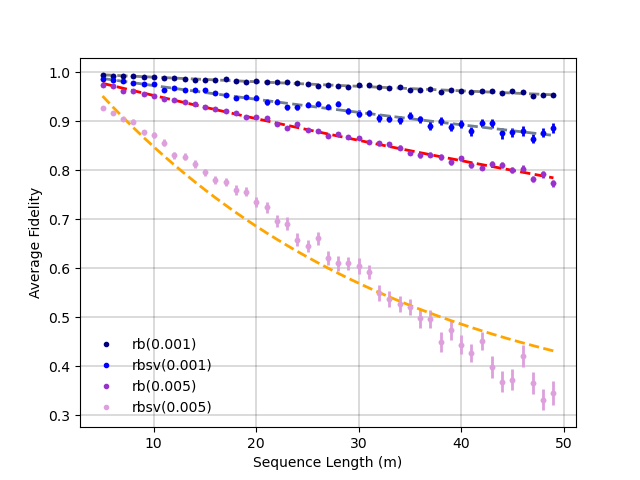}
    }
    \caption{Graphs to show the relationship between the average survival probability estimated from standard RB and the average sequence fidelity estimated from RBSV on a two-qubit system. In Subfig.~\ref{fig: rb_rbsv_0001_001} we plot the results of both protocols for a simple depolarising channel of a) $\varepsilon_i = 0.0001$ and b) $\varepsilon_i = 0.001$ fitting to the standard RB curve: $A_0 + B_0p^m$. In Subfig.~\ref{fig: rb_rbsv_001_005} we plot the results from RB and RBSV for b) $\varepsilon_i = 0.001$ and c) $\varepsilon_i = 0.005$. The parameters $r_{rbsv}$ and $r_{rb}$ returned from both protocols were, for noise strength a) $r_{rbsv} = 0.0003374$, $r_{rb} = 0.0001297$, b) $r_{rbsv} = 0.003752$, $r_{rb} =  0.000898$, c) $r_{rbsv} = 0.024518$, $r_{rb} = 0.004873$.}
    \label{fig:rb_rbsv_graph}
    \end{center}
\end{figure}
In Fig.\ref{fig:rb_rbsv_graph} we present two graphs comparing the results from RBSV and standard RB for a depolarising channel of average strength (a)(top) $\varepsilon_i = 0.0001$, (a)(bottom) and (b)(top) $\varepsilon_i = 0.001$ and (b)(bottom) $\varepsilon_i = 0.005$. We observe that the data from the RBSV protocol for (b)(bottom) $\varepsilon_i = 0.005$ is not as close a fit to the curve as other results, possibly indicating that i) that RBSV works best in a regime of lower noise, and/or ii) that RBSV is more accurate for shorter sequence lengths. The values estimated from RBSV for the average error strength were as follows (a)(top) $r_{rbsv} = 0.0003374$, (a)(bottom) and (b)(top) $r_{rbsv} = 0.003752$ and (b)(bottom) $r_{rbsv} = 0.024518$ and all estimates clearly upper bound the actual average error. An upper bound on the average error rate is a useful figure of merit for any quantum computing device that has an error threshold below which the computations will perform well. If the parameter $r_{rbsv}$ is below this threshold then manufacturers can be confident in their device's performance. 

\subsection{Imperfect Measurements}\label{sec: imperfectmeas}
The RBSV protocol is robust in the presence of preparation errors since each copy does not have to be identical. For the process to work and be robust to SPAM errors, the measurement errors now need to be analysed in the context of stabilizer state verification. Here we define the regime in which RBSV outputs accurate results when these errors occur. 

We write the measurement error in the specific form of \emph{averaged} measurement noise, since this is the parameter extracted from RB.  A noisy measurement of an $n$-qubit stabilizer $s$ is modelled by a noise channel $\Lambda_m$ followed by a perfect measurement. If we assume that the measurement noise is on average depolarising, $\Lambda_m$ is a noise channel representing an independent depolarising noise acting on each qubit touched non-trivially by $s$ with probability $p_{\text{meas}}$. For example, if $s = X_1 \otimes X_2 \otimes 1 ^{\otimes n-2}$, we will denote $|s|$ as the number of qubits touched non-trivially by $s$. In this example $|s| = 2$ since only qubits 1 and 2 are acted on by a non-identity Pauli, and all other qubits are acted on by the Identity. Noiselessly measuring $s$ would correspond to measuring qubits 1 and 2 in the $X$ basis. Under our noise model, with probability $(1-p_{\text{meas}})^{2}$ the measurement is executed noiselessly. With probability $p_{\text{meas}}$ either qubit $1$ or qubit $2$ is affected by an $X$, $Y$ or $Z$ error, and then measured in the $X$ basis. With probability $p_{\text{meas}}^2$ both qubits 1 and 2 are affected by an $X$, $Y$, or $Z$ followed by measuring these two qubits in the $X$ basis. For the SV protocol \cite{MK+20} to work, measurement errors should be very unlikely to occur. The probability for no measurement errors per run of the verification protocol is given by:
\begin{equation}
    P_{\text{perf}}=\prod_{i=1,..,R}(1-p_{\text{meas}})^{|s_i|}\enspace ,
\end{equation}
where $s_i$ is the stabilizer measurement performed on copy $i$. One can easily see that:
\begin{equation}
    P_{\text{perf}} \geq (1-p_{\text{meas}})^{n.R} \enspace .
\end{equation}
In order for the effect of measurement errors to be negligible, we must have
\begin{equation}
\begin{split}
    (1-p_{\text{meas}})^{n\cdot R} &\approx 1\\
    p_{\text{meas}} << \dfrac{1}{n\cdot R} \enspace .
\end{split}
\end{equation}
Thus, for small enough values of $p_{\text{meas}}$ satisfying the above inequality, the contribution of measurement errors can be neglected and therefore the stabilizer state verification techniques developed earlier follow through in this regime.

\subsection{Computational Resources} \label{sec: scaling}
It is important to find the resource cost for implementing RBSV experimentally. Since we find a lower bound on the average sequence fidelity by repeating the same sequence $\tilde{C}_{j,m}$ $N_m$ times to estimating the acceptance probability $\tilde{P}_{\text{acc}_{j,m}}$ we must analyse $N_m$ required to accurately estimate $P_{\text{acc}_{j,m}}$. The actual acceptance probability $P_{\text{acc}_{j,m}}$ itself will be an over-estimate of the survival probability/average fidelity since it could accept on some rounds when the noisy state has the chosen stabilizer in common with the target state. For this reason, we recommend that the number of repetitions used to estimate $P_{\text{acc}_{j,m}}$ be more than the size of the stabilizer group, $N_m \geq 2^n$; however, this is not a rigorous bound and more of a guidance in the NISQ regime for aiming to converge to the actual acceptance probability of some chosen state. Futhermore, as mentioned, RBSV operates in the low error regime and we therefore assume that $P_{\text{acc}_{j,m}}$ is close to 1. The number of copies $R$ resulting in the tightest lower bound was already determined as $R = 1/ln(\frac{1}{P_{\text{acc}_{j,m}}})$ when we assume i.i.d noise between copies. The number $q$ of sequences of different lengths $m$ needs to be large enough such that we include a sufficient spread sequence lengths. To give an idea, a reasonable number of sequences both for experiment and for the method to work is something like $q = 20$. Therefore, the parameters we need to determine the \emph{total} number of experiments (quantum resources) required to perform a typical RBSV protocol and yield meaningful results are $N_m$, $q$, and $K_m$. The sequence length $m$ could also provide a barrier to experimental implementation depending on coherence time of the quantum hardware, and using either very long or very short sequences will over or under-ampify the error channel, it is sufficient to choose a maximum sequence length of $M = 100$. RBSV requires classical computation of the stabilizer group of each random sequence $C_{j,m}$ and therefore we also analyse the classical overhead of our protocol.
\subsubsection*{Quantum overhead}
We use Hoeffding's inequality \cite{H+63} to first determine the number $N_m$ of repetitions required to estimate $P_{\text{acc}_{j,m}}$ to a desired accuracy.
\begin{equation}
    \text{Pr}(|\tilde{P}_{\text{acc}_{j,m}} - P_{\text{acc}_{j,m}}| \geq t) \leq e^{-2N_{m}t^2} \enspace ,
\end{equation}
where we have that $\tilde{P}_{\text{acc}_{j,m}}$ is the empirical mean of the independent random variables bounded by the interval $[0,1]$, since we can treat instances of acceptance as $1$ and of rejection as $0$. For the difference to be minimal $t$ should be very small, $t << 1$ since $P_{\text{acc}_{j,m}} \in [0,1]$ and expected to be very close to 1, and therefore:
\begin{equation}
    N_m \geq \frac{1}{t^2} \enspace ,
\end{equation}
for the probability of the difference being greater than $t$ to be less than or equal to $a \leq e^{-2}$. For a $1 - \delta$ confidence interval of a maximum difference of $\lambda$ between the estimated average sequence fidelity $\tilde{F}_m$ and the actual average sequence fidelity $F_{m}$ the lower bound for the number of random sequences required for standard RB is \cite{WF+14}\cite{HWF+19}: 
\begin{equation}
\label{eq: km}
    K_m = -\frac{\log(2/\delta)}{\log(H(\lambda, \upsilon))} \enspace ,
\end{equation}
where,
\begin{equation}
\label{eq: H}
    H(\lambda, \upsilon) = (\frac{1}{1-\lambda})^{\frac{1-\lambda}{\upsilon + 1}} (\frac{\upsilon}{\upsilon + \lambda})^{\frac{\upsilon + \lambda}{\upsilon + 1}} \enspace ,
\end{equation}
and $\upsilon$ is the variance (due to a finite number of measurements) of the distribution of the samples $\tilde{F}_{j,m}$ used to estimate $\tilde{F}_{m}$. The latest upper bound on this variance for $n$-qubit systems, including SPAM error contribution and unitarity was defined by Helsen et al, see Eq. 11 of \cite{HWF+19}:
\begin{equation}
\label{eq: variance}
\begin{split}
    \upsilon &\leq \frac{(d^2-2)}{4(d-1)^2} r^2 m p^{m-1} \\
    &+ \frac{d^2(1 + 4\eta) r^2}{(d-1)^2} \frac{(m-1)(\frac{p^2}{u})^m - m (\frac{p^2}{u})^{m-1} + 1}{(1 - \frac{p^2}{u})} u^{m-2}\\
    &+ \frac{2\eta d m r}{d-1} p^{m-1} \enspace .
\end{split}
\end{equation}
Where $p$ and $r$ are determined by the fitting procedure in general, and therefore we can treat them as expected values, i.e. we can set a maximum on $r$ and determine $p$ by $p = 1 - \frac{d}{d-1}r$. The parameter $\eta$ encapsulates the SPAM correction and can also be set to an expected value. The unitarity $u = \frac{p^2 + 1}{2}$ and indicates the coherence of the noise, if the noise channel is fully coherent $u = 1$.  A further upper bound, that does not contain SPAM term (Eq.10 of \cite{HWF+19}), is:
\begin{equation}
\label{eq: varianceup}
    \upsilon \leq p^{m-1} \frac{(d^2-1)m}{4(d-1)^2}r^2 + u^{m-2}\frac{d^2m(m-1)}{2(d-1)^2}r^2 \enspace .
\end{equation}
If our lower bound on the average sequence fidelity is tight, which we ensure from our previous assumptions, then we can treat our lower bound as $\tilde{F}_m$ and find that the total number of experiments is lower bounded by:
\begin{equation}
\label{eq: rbsvbound}
    N_{\text{exp}} \geq - \frac{q\log(/\delta)}{t^2 \log(H(\lambda, \upsilon))} \enspace ,
\end{equation}
where this bound includes the confidence in the estimate of the acceptance probability, equivalent to including the confidence in the estimate for the survival probability in standard RB.

\subsubsection*{Classical Overhead}
The only classical computation that must be performed in RBSV is determining the stabilizer group of the ideal final state after running sequence $C_{j,m}$ on an input stabilizer state $\rho$. This will need to be determined for each random sequence $j \in \{K_m\}$ of each sequence $q$ of length $m$. Aaronson and Gottesman introduce the tableau representation in \cite{AG+04} to simulate stabilizer circuits in $\mathcal{O}(n^2)$. Therefore, the classical resources required to perform the RBSV protocol scale as:
\begin{equation}
    N_{\text{class}} = \mathcal{O}(q K_m n^2)
\end{equation}

\section{Background 2.}
\subsection{Interleaved Randomized Benchmarking}
A variant of RB, called interleaved RB~\cite{MGJ+12} (iRB) allows one to characterize the average performance of a specific fixed gate. Standard iRB involves running the Clifford RB protocol as usual, and then running the protocol again with a fixed Clifford gate between each random Clifford, thereby allowing the average error of the fixed Clifford to be extracted from the ratios of the extracted depolarising parameters. The noise is assumed essentially gate and time-independent, as with standard RB, and it is assumed that there is, in general, very low noise on single qubit gates. However, the assumptions of nearly noiseless single qubit gates in standard iRB is not unreasonable, especially given the relatively low fidelities for single qubit gates reported for state-of-art architectures, such as superconducting~\cite{abrams2019implementation} and ion-trap systems~\cite{ballance2016high}. This method could work for any gate-set compatible with standard RB, and an arbitrary fixed gate; however, when the fixed gate is also an element of the gate-set the method is more efficient and scalable as the inverse is far simpler to determine and the assumptions on the noise are more realistic \footnote{For example, if the arbitrary fixed gate involved a complicated and long compilation of native gates then the assumption of essentially gate and time-independent noise would be extremely unrealistic.}. 
\par 
We denote the noise channel of the fixed Clifford gate as $\Lambda_{C}$ and the average noise channel of all random Clifford elements as $\Lambda$. The combined noise channel during the second run is $\Lambda_{\overline{C}} = \Lambda \circ \Lambda_{C}$, it is simple therefore to estimate the RB parameter for the fixed Clifford gate, $r_{C}^{est}$: 
\begin{equation}
    r_{C}^{est} = (d - 1) ( 1 - \frac{p_{\overline{C}}}{p})/d \enspace ,
\end{equation}
where $d = 2^n$ is the dimension of the Hilbert space, $p_{\overline{C}}$ is the parameter extracted from the second run, where the fixed gate has been added, and $p$ is the parameter extracted from the first standard RB run. The actual average gate infidelity for the fixed Clifford gate, $r_{C}$ lies within the interval $|r_{C}^{est} - E, r_{C}^{est} + E|$, where $E$ can be thought of as an error arising from non-uniformly random gates, see Eq.5 in\cite{MGJ+12}. Since $p_{\overline{C}}$ characterises the mixed noise channel $\Lambda_{\overline{C}}$ and $p$ characterises the noise channel $\Lambda$, the characterisation of $\Lambda_{C}$ is simply $p_{\overline{C}}/p$.   

\subsection{Gate Synthesis of Clifford Circuits}\label{sec: GSinRB}
RB, as previously mentioned, usually provides a characterisation of the average error per Clifford element (of the full Clifford group) on a quantum device. It is crucial to include a treatment of non-Clifford gates in the benchmark, especially if those non-Clifford gates are native to the quantum device in question. Moreover, a single non-Clifford gate together with the ability to perform all Clifford gates (e.g. via a generating set of Cliffords) form a universal gate-set. Restricting algorithms to specific gate-sets is known as gate synthesis \cite{MDM+13}, and is used to compile other quantum gates from the elements of this gate-set, such that the overall computation is still implemented with the original gate-set.
\par 
Gate synthesis is most often applied to find optimal compositions of universal gate-sets to form interesting quantum operators. The Solovay-Kitaev (SK) theorem \cite{skth} shows that, given a dense (approximately universal) subset $\mathcal{U}$ of $U(d)$ for a fixed dimension $d$, approximating any unitary $U \in U(d)$ up to some $\varepsilon$ can be done efficiently using sequences of gates from $\mathcal{U}$ whose length scales poly-logarithmically with $1/\varepsilon$. Up until very recently,  $\mathcal{U}$ was required to be symmetric (contain the inverses of the unitaries), although some partial relaxations had been explored \cite{bouland2017trading}.  However, recently \cite{epsilon-nets} it was shown that this constraint can be relaxed. This largely opens up the sets $\mathcal{U}$ that can be used as native gates in a quantum hardware. In this work we are focused on the synthesis of the Clifford group, i.e. instead of decomposing an arbitrary gate into a known gate-set (e.g. Clifford+T \cite{BK06} or CNOT-dihedral group \cite{garion2020structure}) we are interested in decomposing the Clifford group (or generators thereof) into a combination of Clifford and non-Clifford gates; with a preference on the non-Cliffords being native gates of the quantum hardware. The motivation here is to determine a reliable estimate for the error behaviour of non-Clifford gates when they are part of a circuit composed of Clifford and non-Clifford gates, \emph{and} how such a circuit will perform on a chosen quantum device, on average.

\section{Interleaved RB + Gate Synthesis}\label{sec: iRB+GS}
We present a modification to iRB by replacing the fixed Clifford gate with its synthesised counterpart made up of two non-Clifford native gates and single qubit Clifford gates. Since Clifford elements are composed of a number of single and multi-qubit gates we inherent the assumption of essentially noiseless single qubit gates from \cite{MGJ+12} in order to isolate the average error of a fixed multi-qubit gate. In Sec.~\ref{sec: construction} we present a specific example, and in Sec.~\ref{sec: rotations} we extend this example to a family of non-Cliffords allowing for more than two non-Cliffords in the synthesis. 

We define the average fidelity of a noise channel $\Lambda$ as
$\overline{F}_{\Lambda} \equiv \overline{F}_{\Lambda, \mathcal{I}}$, and $\Lambda_d$ is the unique depolarising channel of $\Lambda$ with the same average fidelity. As before, we have that $\Lambda_{\overline{C}}$ represents the average noise channel of the fixed Clifford element $C$. We denote the native non-Clifford gate to be benchmarked as $N$ and the noise channel of $N$ as $\Lambda_N$.

Similarly to iRB, we first run the standard RB protocol for random Cliffords and extract the depolarising parameter $p$, where the corresponding average fidelity $\overline{F}_{\Lambda}$ is equal to $1 - (d-1)(1-p)/d$. In the second run, we include our synthesised $C$ gate, with the combined noise channel: $\Lambda_{\overline{C}} \equiv \Lambda_N \circ \Lambda_N \circ \Lambda$. We extract the parameter $p_{\overline{C}}$ corresponding to the mixed error channel, and the corresponding average fidelity $\overline{F}_{\Lambda_{\overline{C}}}$ is equal to $1 - (d-1)(1 - p_{\overline{C}})/d$.

We make the assumption that the noise channels for each non-Clifford gate $\Lambda_N$ are on average the same and compose as a product of channels $\Lambda_N \circ \Lambda_N \equiv (\Lambda_N)^2$. We can therefore determine our estimated error rate (RB parameter) $r_N^{est}$ for the native $N$ gate as: 
\begin{equation}
    r_N^{est} = \frac{d-1}{d} (1 - \sqrt{\frac{p_{\overline{C}}}{p}}) \enspace ,
\end{equation}
since we have that $p_{\overline{C}}$ characterises $(\Lambda_N)^2 \circ \Lambda$ and $p$ characterises $\Lambda$, therefore $\frac{p_{\overline{C}}}{p}$ characterises $\frac{(\Lambda_N)^2 \Lambda}{\Lambda}$ and therefore $\sqrt{\frac{p_{\overline{C}}}{p}}$ characterises $\sqrt{(\Lambda_N)^2} = \Lambda_N$. Using the results of \cite{MGJ+12} we find  the upper bound between our estimate $r_N^{est}$ and the actual average gate infidelity: $r_N = 1 - \overline{F}_{\Lambda_N}$. We extend this for three different noise models in the following section.

With this simple modification, iRB+GS provides a way to benchmark any non-Clifford or native gate that can compose with other Clifford gates such that the overall element is Clifford.

\subsection{Noise analysis} \label{sec: noiseanalysis}
We analyse the error in our estimate $r_N^{est}$ that arises from approximately twirling, i.e. due to not having perfectly random gates, the noise of the native gates together with the average Clifford noise. We explore three different noise channels for the native non-Clifford gate $N$ : a depolarising channel, a $\delta$-depolarising channel and a Pauli noise channel. We detail the proof for the first noise channel, which uses arguments from \cite{MGJ+12} and state the results of the other two, with their proofs in App.~\ref{app: GSproof}. 
\par
We first define the noise channel for the combination of non-Clifford and Clifford gates when the noise on all gates has been twirled: $\Pi_{\overline{C}} = (\Lambda_{N_d})^2 \circ \Lambda_d$. 
\par 
We know that the twirled noise channels are equal to a depolarising channel, and therefore we can quantify the difference between the fidelity of the channel $\Lambda_{\overline{C}}$ and $\Pi_{\overline{C}}$. 
\begin{equation}
\begin{split}
    | \overline{F}_{\Lambda_{\overline{C}}} - \overline{F}_{\Pi_{\overline{C}}} | &= | p_{\overline{C}} + \frac{1-p_{\overline{C}}}{d}  - p p^2_{N} - \frac{1- p p_{N}^2}{d} | \\ 
    | \overline{F}_{\Lambda_{\overline{C}}} - \overline{F}_{\Pi_{\overline{C}}} | &= \frac{d-1}{d} | p_{\overline{C}} - p p^2_{N} |\enspace .
    \label{eq: proof}
\end{split}
\end{equation}
where we have used the fact that the twirled channels $\Pi_{\overline{C}} = (\Lambda_{N_d})^2 \circ \Lambda_d$ become depolarising channels, where $p_N$ characterises the depolarising parameter of channel $\Lambda_N$ and $p$ characterises the depolarising channel $\Lambda_d$. Variable $p_{\overline{C}}$ is the depolarising parameter for the combined noise channel $\Lambda_{\overline{C}}$.

When $\Lambda_N$ is itself a depolarising channel, the difference between the estimated error rate $r_N^{est}$ and the actual average error $r_N$ is bounded as follows:
\begin{equation}\label{eq: varepsilon1}
    |r_N^{est} -  r_N| \leq \sqrt{\frac{d-1}{d} \frac{E'}{p}} \enspace ,
\end{equation}
\par 
where $E ' = \frac{2(d^2 -1)(1-p)}{d^2} + 4 \sqrt{1-p} \sqrt{d^2-1}$

\begin{proof}
The left hand side of Eq.~\ref{eq: proof} is upper bounded as:
\begin{equation}
\begin{split}
|\overline{F}_{\Lambda_{\overline{C}}} - F_{\Pi_{\overline{C}}}| &\leq ||\Lambda_N - \Lambda_{N_d}||_{\diamond}\\
&+ ||\Lambda_N - \Lambda_{N_d}||_{\diamond} + ||\Lambda - \Lambda_d||_{\diamond}\\
&\leq ||\Lambda - \Lambda_d||_{\diamond}
\end{split}
\end{equation}
with $|| \cdot ||_{\diamond}$ the diamond norm distance \cite{K+97}. Which, stating the proof from the original iRB \cite{MGJ+12}, we can re-write as:
\begin{equation}
\begin{split}
    |\overline{F}_{\Lambda_{\overline{C}}} - \overline{F}_{\Pi_{\overline{C}}}| &\leq ||\Lambda - \mathcal{I}||_{\diamond} + ||\Lambda_d - \mathcal{I}||_{\diamond}\\
    |\overline{F}_{\Lambda_{\overline{C}}} - F_{\Pi_{\overline{C}}}| &\leq 4 \sqrt{1 - p}\sqrt{d^2 - 1} \\
    &+ \frac{2(d^2 - 1)(1-p)}{d^2} = E'
\end{split}
\end{equation}

by the triangle inequality and using the fact that $||\Lambda_d - \mathcal{I}||_{\diamond} = \frac{2(d^2 - 1)(1-p)}{d^2}$\cite{MGE+12}, and for arbitrary $\Lambda$, $||\Lambda - \mathcal{I}||_{\diamond} \leq 4 \sqrt{1 - p}\sqrt{d^2 - 1}$\cite{SR+11}.

All the quantities in  Eq.~\ref{eq: proof} are now known from the experiment or are bounded. Our estimate of the depolarising parameter for the noise channel $\Lambda_N$ is $p^{est}_N = \sqrt{\frac{p_{\overline{C}}}{p}}$ and we can now re-arrange Eq~\ref{eq: proof} such that we have:

\begin{equation}
    |  p^2_N - p^{est \text{ } 2}_N  | \leq \frac{d}{d-1} \frac{E'}{p} 
\end{equation}
Then, finding the root we have:
\begin{equation}
    |p_{N} - p_{N}^{est}| \leq \sqrt{\frac{d}{d-1}\frac{E'}{p}} \enspace ,
\end{equation}
and since we know that for any noise channel $\Lambda$, $r = \frac{d}{d-1}p$, we can re-arrange:
\begin{equation}
\begin{split}
    | \frac{d-1}{d}( (1- p_N) - (1- p_N^{est})) | &\leq  \frac{d-1}{d} \sqrt{\frac{d}{d-1} \frac{E'}{p}}   \\
    | r_N - r_N^{est} | &\leq  \sqrt{\frac{d-1}{d} \frac{E'}{p}} \enspace .
\end{split}
\end{equation}
which gives us Eq.~\ref{eq: varepsilon1}.
\end{proof}

Using similar arguments, we can see that if the noise channel $\Lambda_N$ is \emph{$\delta-$depolarising} for a known upper bound $\delta$ then:
\begin{equation}
\label{eq_delta}
| r_N - r_N^{est} | \leq \sqrt{\frac{d-1}{d} \frac{E' + 2 \delta}{p}} \enspace .
\end{equation}

And in the case that $\Lambda_N$ is Pauli noise:
\begin{equation}
\label{eq: pauli}
    | r_N - r_N^{est} | \leq  \sqrt{\frac{d-1}{d} \frac{E''}{p}}
\end{equation}

where  $E''=  \frac{6(d^2 -1)(1-p)}{d^2} + 4 \sqrt{1-p} \sqrt{d^2-1}$

Proofs of Eq.~\ref{eq_delta} and Eq.~\ref{eq: pauli} can be found in App.~\ref{app: GSproof}.

\subsection{Synthesis of two-qubit Clifford generating group}\label{sec: construction}
We present a construction for a \emph{two-qubit} system. We aim to include a non-Clifford gate that is native to a chosen hardware in our two-qubit Clifford circuit, in order for it to be included in the benchmark. First, we need a description of the two-qubit Clifford gate(s) that we want to synthesise (with the non-Clifford gate). We require that each two-qubit Clifford gate that we synthesise use a fixed number of instances of the native non-Clifford gate. It is also required that the synthesis of our two-qubit Clifford gates must be non-trivial, i.e. it should not be implementable by the exact same Hamiltonian as the original (non-synthesised) Clifford gate\footnote{For example, a rotation around $z$-axis of $90$ degrees by two separate rotations  around $z$-axis of $45$ degrees is effectively the same Hamiltonian.}. Following, we demonstrate a brute-force synthesis construction, however an open question that comes out of our paper is how to modify existing elaborate synthesis techniques to the case of synthesising Cliffords from a gate-set that includes non-Clifford gates.\par
\begin{figure}[h!]
\begin{center}
\begin{tabular}{|l|p{70mm}|}
     \hline 
     $\mathcal{C}_{\text{gen}_2}$ & synthesised $C_{\text{gen}_{2}}$ \\[1.5ex]
     \hline
     $I_i \otimes P_j$ & $(X_i \otimes I_j)CP_{ij}(X_i \otimes I_j) CP_{ij}$ \\[2ex]
     \hline
     $P_i \otimes P_j$ & $CP_{ij} (X_i \otimes X_j) CP_{ij}^{\dagger}(X_i \otimes X_j)$\\[2ex]
     \hline 
     $CNOT_{ij}$ & $(P_i^{\dagger} \otimes H_j X_j) CP_{ij} (I_i \otimes X_j) CP_{ij}^{\dagger} (I_i \otimes H_j)$\\[2ex]
     \hline
     $I_i \otimes H_j$ & $(P_{i}^{\dagger} \otimes I_j) CP_{ij} (I_i \otimes X_j) CP_{ij} (I_i \otimes H_j)$ \\[2ex]
     \hline 
     $H_i \otimes H_j$ & $(P_i^{\dagger} \otimes I_j) CP_{ij} (I_i \otimes X_j) CP_{ij} (H_i \otimes H_j)$\\[2ex]
     \hline 
     $I_i \otimes P_j^{\dagger}$ & $(X_i \otimes I) CP_{ij}^{\dagger} (X_i \otimes I_j) CP_{ij}^{\dagger}$ \\ [2ex]
     \hline 
     $P_i^{\dagger} \otimes P_j^{\dagger}$ & $CP_{ij}^{\dagger} (X_i \otimes X_j) CP_{ij}(X_i \otimes X_j)$\\[2ex]
     \hline 
\end{tabular}
\end{center}
\caption{The construction for synthesising the symmetric two-qubit Clifford group generators with two controlled-Phase gates $CP$ synthesised in each element. The indices $i, j$ indicate which qubit each gate is acting on, for example $CP_{ij}$ indicates that the $i_{th}$ qubit is the control qubit, and $j_{th}$ qubit is the target qubit, etc.}
\label{fig: synthesistable}
\end{figure}
The controlled-Phase (CP) gate fits all of the above requirements, where we assume that its conjugate $CP^{\dagger}$ has equivalent noise behaviour. We find the synthesis of the self-inverting generator set of the two-qubit Clifford group, whereby self-inverting means a symmetric set of operators such that the inverses of each operator is contained in the set. We call this set $\mathcal{C}_{\text{gen}}$ and define: $\mathcal{C}_{\text{gen}} = \{H, P, P^{\dagger}, CNOT\}$. Where $H = \frac{1}{\sqrt{2}}\begin{psmallmatrix}
1 & 1 \\
1 & -1  
\end{psmallmatrix}$ is the single qubit Hadamard gate, $P= \begin{psmallmatrix}
1 & 0 \\
0 & i  
\end{psmallmatrix}$ is the single qubit phase-gate, and because $P$ is Unitary, $P^{-1} = P^{\dagger}$, and $CNOT= \begin{psmallmatrix}
I & 0  \\
0 & X  
\end{psmallmatrix}$ is the two qubit controlled-X gate, where $X=\begin{psmallmatrix}
0 & 1  \\
1 & 0  
\end{psmallmatrix}$.

In Fig.~\ref{fig: synthesistable}, there are two $CP$ or $CP^{\dagger}$ gates in each Clifford generator. This is because when we are estimating the average error (using iRB+GS) of each individual $CP$ gate we assume that the error composes as a product of the errors on each gate and this assumption is less realistic with more than a few gates, so having a minimal (in this case, two) number of non-Cliffords is desirable.

\subsection{Case study: Family of Rotations}\label{sec: rotations}
We generalise the earlier construction for a wider family of non-Clifford gates by considering the family of controlled-$P(k)$ gates, where :
\begin{equation}
    P(k) = 
    \begin{psmallmatrix}
    1 & 0 \\
    0 & e^{i 2 \pi /2^k} 
    \end{psmallmatrix} \enspace ,
\end{equation}
where $k$ is a positive integer, and in the case of the standard controlled-Phase gate from the previous construction $k=2$. Members of this family are used in various quantum algorithms such as quantum fourier transformation. As an example, we can synthesise the generator $I \otimes P(k=2)$, chosen from our construction (for two-qubits) in Fig.~\ref{fig: synthesistable} such that it contains two $CP(k=3)$ gates: 
\begin{equation}
\begin{split}
I \otimes P(2) &= (I \otimes P(3) )(I \otimes P(3))\\
=&(X \otimes I) CP(3) (X \otimes I) CP(3)\\
&(X \otimes I) CP(3) (X \otimes I) CP(3) 
\end{split} \enspace .
\end{equation}
Where we notice that:
\begin{equation}
\label{eq: pk}
    P(k-1) = P(k)P(k) \enspace ,
\end{equation}
and:
\begin{equation}
\label{eq: ipk}
   I \otimes P(k-1) = (X \otimes I) CP(k) (X \otimes I) CP(k) \enspace .
\end{equation}

Now we set our fixed Clifford gate to $C = I \otimes P(2)$ to analyse iRB+GS for this family of gates. Using the above Eqs.~\ref{eq: pk} and \ref{eq: ipk}, we can determine that synthesising this generator for general $k$ involves $2^{k-1}$ identical non-Clifford $N = CP(k)$ gates, where we assume on average the same noise for each. We can therefore derive a new estimated RB parameter as:
\begin{equation}
    r_{N}^{est} = \frac{d-1}{d}(1 - 2^{k-1} \sqrt{\frac{p_{\overline{C}}}{p}}) \enspace ,
\end{equation}
where we have used the fact that we can rewrite Eq.~\ref{eq: proof} for general $k$ as:
\begin{eqnarray}
    | \overline{F}_{\Lambda_{\overline{C}}} - \overline{F}_{\Pi_{\overline{C}}} | = \frac{d-1}{d} | p_{\overline{C}} - p p^{k-1}_{N} |
\label{eq_important}
\end{eqnarray}
And therefore, for a depolarising noise channel $\Lambda_N$ on each $N = CP(k)$ gate, we would have: 
\begin{equation}
    |  p^{k-1}_N - p^{est \text{ } k-1}_N  | \leq \frac{d}{d-1} \frac{E'}{p} \enspace ,
\end{equation}
and, therefore the difference between the actual RB parameter for $N$ gates and our estimated one is upper bounded by:
\begin{eqnarray}
    | r_N - r_N^{est} | \leq  \sqrt{k-1}{\frac{d-1}{d} \frac{E'}{p}} \enspace ,
\end{eqnarray}
with $E'$ as specified earlier in Sec.~\ref{sec: noiseanalysis}. Similarly, the error on the estimate for the $\delta$-depolarising and Pauli noise channels can be trivially derived, as before.

\section{Application of Methods: Generator RB}
Most RB protocols suffer from the same physically limiting problem: the amount of resources required to perform them is incredibly large for system sizes of more than a few qubits. This is due to the fact that most multi-qubit RB schemes require one to implement the Clifford group, which means that each $n$-qubit Clifford element must be decomposed with the native gates of the system; for $n > 2$ this introduces a significant overhead that becomes quickly impractical. In fact, a five qubit system is the largest system size benchmarked (to date) by direct RB \cite{PCR+19} which actually utilises the generators of the Clifford group $\mathcal{C}_n$. For standard RB the largest system size to be benchmarked, to date, is a three qubit system \cite{mckay2019three}. Therefore, it is key to be able to benchmark gates requiring fewer native gates to implement, and subsequently shorter sequences, that may be more efficiently implemented on physical hardware. Fran\c{c}a and Hashagen \cite{FH+18} introduce another method for RB with generators which we will call generator RB.
\par 
To simplify the explanation of their method we discuss it directly in the context of the Clifford group since our framework dictates that our gates must be Clifford. As mentioned in Sec.~\ref{sec: construction} the gates that generate the Clifford group are the Hadamard gate, the Phase-gate and the CNOT gate. In order for generator RB to work, the set of generators must be closed under inversion, and so as used for our synthesis construction, we have:
\begin{equation}
    \mathcal{C}_{\text{gen}} = \{H_i, P_i, P_i^{\dagger}, CNOT_{ij}\}
    \nonumber
\end{equation}
In \cite{FH+18} it is shown that randomized benchmarking the generators of a group $\mathcal{G}$ is feasible if the set of generators is closed under inversion and \emph{rapidly mixing}. Where the mixing time denotes how quickly a chain of randomly applied generators converge to the Haar measure on the group $\mathcal{G}$. The authors conjecture (based on their numerics and the results of \cite{41}) that the above set $\mathcal{C}_{\text{gen}}$ is rapidly mixing with $t_{\text{mix}} = O (n^2 \log(n))$, i.e. that after this time $t_{\text{mix}}$ a set of generators will converge to the Clifford $2$-design. Where $t_{\text{mix}}$, in this case, indicates the number of generator gates applied in order to reach convergence. 

Therefore, it is clear that if we pick gates uniformly at random from $\mathcal{C}_{\text{gen}}$, i.e. $C_{g1} C_{g2} \cdots $, $C_{gi} \in \mathcal{C}_{\text{gen}}$ it follows that $C_{gb} C_{gb-1} \cdots C_{g1}$ will be approximately distributed like the Haar measure on $\mathcal{C}_n$ when $b \simeq t_{\text{mix}}$. Once the sequence length of random generators passes the threshold $b$, the RB method works as usual and the fitting procedure is enacted after this point to find the figure of merit per generator rather than full group element. Generator RB \cite{FH+18} provides reliable results only in the low error regime, i.e. with high fidelity combinations of gates/qubits and specifically for $\delta$-covariant noise channels, particularly when $\delta = p$ is close to $1$. This noise regime is highly compatible with the RBSV protocol as our results are most accurate under these constraints. Moreover, $b \simeq t_{\text{mix}}$ is not an impractical amount of generators and in fact in \cite{FH+18} they found that for $b = 10$ generator gates, and $p$ close to $1$ they could determine the average fidelity of the noise channel within $10^{-3}$ of the exact average fidelity on a (simulated) five qubit system; results being virtually indistinguishable from RB for the exact same noise channel.

The drawbacks of generator RB are that it requires an arbitrary group element to be applied after the sequence of randomly mixing generators to invert the preceding sequence. Though this is not infeasible for $\mathcal{C}_n$, it does mean that the same noise assumption on the inverse step made in RB is made here, i.e. there is a single error associated with the inverse step and it is the same on average as the noise on the generators in the sequence. Since this method was developed to benchmark generators and not full group elements it is unfortunate to have a limiting assumption such as this. Our RBSV protocol does not require an inverse step and is directly applicable to generator RB, offering a way to perform the benchmark without the arbitrary group element. RBSV could also be used in conjunction with direct RB \cite{PCR+19} in place of the Clifford measurement. It is worth noting that RBSV works for any sequences of gates that form a stabilizer circuit, and therefore RBSV is compatible with any generating set of gates for $\mathcal{C}_n$.

Using generator gates rather than the full Clifford group also means that a smaller number of native gates is required to implement the gate being benchmarked. For iRB+GS we require a low number of non-Clifford gates to be synthesised into the fixed Clifford element. We expect that applying iRB+GS with generators rather than the full Clifford group would make the process easier and more realistic. Generators are sufficient to allow for a non-trivial amount of non-Clifford gates in the computation, whilst the quantum computational resources required to run the sequences will be lower which is compatible with iRB+GS since the second run of the protocol requires double the sequence length of gates. Generator RB provides a way to reduce the resource overhead of the method making iRB+GS more practical.

\section{Conclusion}
In summary, we have modified the standard Clifford RB technique of \cite{MGE+12,MGE+11} in two ways. First, we remove the inverse requirement in RB by using the protocol of SV \cite{MK+20}, a modification we name RBSV. One advantage of RBSV is that it makes existing schemes of RB tailored to the NISQ  era \cite{FH+18,PCR+19} more practical. Second, we incorporate gate synthesis in the usual Clifford iRB \cite{MGJ+12} and show that this can be used to characterize new sets of non-Clifford gates, albeit with some assumptions on the noise channels.

The RBSV technique works most reliably when the noise rates affecting gates, preparations, and measurements are low, as evidenced by our analytical and numerical calculations. This is due in part to the low robustness to noise of the schemes of quantum verification \cite{GKK19} (which RBSV is based upon). An interesting question is whether techniques of robust quantum verification recently developed in \cite{kashefi2020securing} could be used to modify RBSV to provide reliable results in regimes where the noise is not very low.

Another direction to pursue would be making our gate synthesis and iRB procedure more systematic. Indeed, rather than the brute force expansion technique of a Clifford gate in terms of specific non-Cliffords used here, one could think of developing a recipe for expansion which could potentially incorporate a wide variety of non-Cliffords. This could aid in making our techniques easily applicable to different hardware, each with their own sets of native non-Clifford gates.

\section{Acknowledgements}
We thank Damian Markham for helpful discussions. ED acknowledges support from the Doctoral Training Partnership (EP/N509711/1) under project No.1951737. EK acknowledges support from the following:  EPSRC Verification of Quantum Technology grant (EP/N003829/1) and UK Quantum Technology Hub: NQIT grant (EP/M013243/1) and the EU Flagship Quantum Internet Alliance (QIA) project. RM and EK acknowledge support from the grant: Innovate UK Commercialising Quantum Technologies (application number: 44167).

\bibliographystyle{apsrev}
\bibliography{Bibliography.bib}
\appendix 
\section{Standard Randomized Benchmarking}\label{app: RB}
Randomized benchmarking  (RB) is an experimental method that, at it's core, provides an efficient way to determine the \emph{average} performance of a gate-set on a specific quantum device. There have been many developments since the seminal RB literature, and we discuss those most relevant to our work in. Standard RB protocols consist of running many random sequences (of length $m$) of unitary gates, uniformly sampled from a gate-set $\mathcal{U}$, on an $n$-qubit device such that if the gates were implemented perfectly the device would return to its initial state. The probability of the device returning to its initial state is measured and recorded as the \emph{average survival probability} of sequence length $m$. This procedure is repeated for various lengths $m \in \{M\}$. The resulting average survival probabilities are plotted against the sequence length and fit to an exponential decay curve derived from RB theory, from which the average fidelity and therefore average error rate $r$ of the gates may be found. 
\\
\par 
Crucial to randomized benchmarking is the concept of \emph{twirling}, essentially averaging over random unitary conjugations. Twirling a quantum channel $\Lambda$ over the (infinite) set of unitaries distributed according to the Haar measure $\mu$ over the entire unitary space $U(d)$, produces a \emph{depolarising channel}; where $d = 2^n$ is the dimension of the $n$-qubit Hilbert space and the Haar measure is a measure of uniformity. A depolarising channel is a simple error channel that may be easily characterised. If we set $\Lambda$ to be an arbitrary quantum channel acting on a state $\rho$, then the Haar twirled channel $\Lambda_t (\rho)$ is as follows:
\begin{equation}
\label{eq: twirl}
    \Lambda_t(\rho) = \int_{U(d)} U^{\dagger} \Lambda(U\rho U^{\dagger})U d\mu = p\rho + (1-p)\frac{\mathcal{I}}{d} \enspace ,
\end{equation}
where the integral indicates the process of twirling and the right-hand side is the depolarising channel; the state $\rho$ remains unchanged under this process with probability $p$ and otherwise returns the maximally mixed state, $\frac{\mathcal{I}}{d}$. In this way, twirling can be understood as a means to view the average effect of \emph{any} noise channel as a depolarising noise, thereby making it easier to study the average performance of noisy gates. \\ \par
Each gate that we are interested in characterising in RB, is chosen from the gate-set $\mathcal{U} \in U(d)$, according to some probability measure $\tau$ with support over $\mathcal{U}$. The couple $\{\tau, \mathcal{U}\}$ commonly forms an exact \emph{unitary 2-design}, i.e. the average, over the choice of $U$, of \emph{any} polynomial $P_{(2,2)}(U)$ of degree 2 (or less) in the matrix elements of $U \in U(d)$, and degree 2 (or less) in the matrix elements of $U^{\dagger}$ is \emph{exactly} the same when $U$ is either chosen from the Haar measure on $U(d)$ or from $\{\tau, \mathcal{U}\}$. In Eq.~\ref{eq: twirl}, $U^{\dagger} \Lambda (U \rho U^{\dagger}) U$ is a $P_{(2,2)}(U)$ polynomial, and therefore:
\begin{equation}
    \Lambda_t(\rho) = \int_{U(d)} P_{(2,2)} (U) d\mu = \int_{\mathcal{U}} P_{(2,2)} (U) d\tau \enspace ,
    \label{eq: t-design-twirl}
\end{equation}
where the right-hand side is the twirl over the couple $\{\tau, \mathcal{U}\}$. It is clear to see that twirling a quantum channel over a unitary 2-design produces the same output as twirling over the Haar twirl. Since $\mathcal{U}$ is a \emph{finite} set of unitaries, the integral may be written as a sum and we have: 
\begin{equation}
\label{eq: t-design-dep}
\frac{1}{|\mathcal{U}|} \sum_{j=1}^{|\mathcal{U}|} P_{(2,2)}(U_j) = p\rho + (1-p) \frac{\mathcal{I}}{d} \enspace .
\end{equation}
A well known example of an exact unitary 2-design is the Clifford group on $n$-qubits $C_{d}$, as well as some sub-groups of $C_{d}$. Indeed, multiple RB protocols have been developed to test the performance of Clifford unitaries; although RB itself is not limited only to Clifford unitaries, and has been used to characterise the performance of so-called $T$-gates which, together with Clifford gates, suffice for universal quantum computation. Furthermore, several approaches to modifying RB to work beyond the exact 2-design condition have been considered. 
\\ \par 
As mentioned, in RB we want to find a measure of the \emph{average} performance of a set of quantum gates. To do this we look at the fidelity of the imperfectly implemented quantum gates with their perfect counterparts, where fidelity is a measure quantifying the distance between the outputs of $\Lambda_U$ and $U$ acting on a pure state $\rho = \ket{\psi}\bra{\psi}$. This can be equated to the fidelity between the noise channel $\Lambda_\varepsilon$ and the Identity $\mathcal{I}$: $F(\Lambda_U(\rho), U(\rho)) = F(\Lambda_\varepsilon(\rho), \mathcal{I}(\rho)) = Tr(\rho \ \Lambda_\varepsilon(\rho))$; where $\Lambda_\varepsilon = U^{\dagger} \circ \Lambda_U$, therefore if $\Lambda_\varepsilon = \mathcal{I}$, $\Lambda_U = U$. We are interested in the \emph{average} fidelity of a set of quantum gates, which is the fidelity of the error channel $\Lambda_\varepsilon$ (wrt to $\mathcal{I}$) averaged over all pure input states: $\overline{F(\Lambda_U, U)} = \overline{F(\Lambda_\varepsilon, \mathcal{I})}$. 
\\ \par 
The goal then of RB is to mimic the above twirl (Eq.~\ref{eq: t-design-dep}) over the exact 2-design $\{\tau,\mathcal{U}\}$ in order to simplify the gates' average error channel $\overline{\Lambda_\varepsilon}$ into a depolarising channel to extract the parameter $p$. 
Under this twirl, the average fidelity of the noise channel $\Lambda_\varepsilon$ with respect to the Identity $\mathcal{I}$ is invariant:
\begin{equation}
    \overline{F(\Lambda_\varepsilon, \mathcal{I})} = \overline{F(\Lambda_{\varepsilon, t}, \mathcal{I})} = p + \frac{1-p}{d} \enspace .
\end{equation}
Where $\Lambda_{\varepsilon ,t}$ is the noise channel twirled. In RB, we mimic this twirl by finding the average survival probability ($f_m$) for sequence lengths $m$, which converges (with appropriately large repetitions, number of random permutations of gates and sequence lengths) to the (uniformly) average sequence fidelity $\overline{F}_{\text{seq}_m}$ which is equal to $p^m + \frac{(1 - p^m)}{d}$ when there are no errors from state preparation and measurement (SPAM). 
\\ \par 
One of the properties of RB is that these so-called SPAM errors can be characterised. 
The SPAM errors are factored into the RB data fitting coefficients, and when noise is assumed gate and time-independent the RB data can be fit to the following curve: $F^m(p) = A_0 + B_0 p^{m}$, where $A_0$ and $B_0$ contain SPAM contributions and $p$ can be extracted. The average fidelity and the average error rate $r = 1 - \overline{F}_{\text{gate}}$ for each gate may then be calculated from the depolarising parameter $p$. 
\subsection{Average Fidelity and Survival Probability}
We explicitly derive the relationship between the average sequence fidelity and the survival probability of randomized benchmarking schemes, where the two terms are often used interchangeably. 
The average fidelity of a quantum process $U_{m}$ on an initial state $\rho$, can be written explicitly as: $\overline{F}_{m} = F(\Lambda_{m} \circ U_m \circ \cdots \Lambda_1 \circ U_1 \circ \Lambda_{\text{prep}}(\rho), U_m \cdots U_1 \rho (U_m \cdots U_1)^{\dagger})$. Where the imperfectly implemented gate is $\Lambda_m \circ U_m$ and $\Lambda_{\text{prep}}(\rho)$ indicates an imperfectly prepared initial state. We can rewrite this expression as:
\begin{equation}
\begin{split} 
    \overline{F}_{m} &= Tr(\ket{\psi_m}\bra{\psi_m} \Lambda_m \circ U_m \\
    &\cdots \Lambda_1 \circ U_1 \circ \Lambda_{\text{prep}} (\rho)) , \\
    \text{where} \ \ket{\psi_m} &= U_m \cdots U_1 \ket{0_n} \enspace .
\end{split}
\end{equation}
Now, letting $U = U_1 \cdots U_m$, and setting the ideal initial state to $\rho = (\ket{0}\bra{0})^{\otimes n} := \ket{0_n}\bra{0_n}$, we can rewrite the above as:
\begin{multline}
\label{eqfidelity2}
  \overline{F}_m =Tr(\ket{\psi_m}\bra{\psi_m} \Lambda_m \circ U_m \circ \cdots \Lambda_1 \circ U_1 \circ  \Lambda_{prep}(\rho))\\
  =Tr(U\ket{0_n}\bra{0_n}U^{\dagger} UU^{\dagger}\Lambda_m \circ U_m \circ \cdots \Lambda_1 \circ U_1 \circ \Lambda_{prep}(\rho)U^{\dagger}U)\\ =Tr(\ket{0_n}\bra{0_n}U^{\dagger}\Lambda_m \circ U_m \circ \cdots \Lambda_1 \circ U_1 \circ \Lambda_{prep}(\rho)U)\\=Tr(\ket{0_n}\bra{0_n}U^{\dagger} \circ \Lambda_m \circ U_m \circ \cdots \Lambda_1 \circ U_1 \circ \Lambda_{prep}(\rho)).
\end{multline}
When measurements are assumed perfect, the rightmost part of Eq.~\ref{eqfidelity2} is exactly equal to the survival probability $f_m$, assuming a perfect inversion step. That is, the exact same analysis as in \cite{MGE+12} can be performed, while taking the noise channel on the inverse gate to be $\Lambda_{U^{\dagger}}=\mathbf{1}$. From this analysis it can be seen that the average sequence fidelity $\overline{F}_m$ will follow the fitted curve $F^m(p)$ corresponding to  twirling the $m$ gates $U_1,...,U_m$ of the RB circuit. 

\section{Proofs for Section V.}\label{app: GSproof}
We can model the $\delta$-depolarising noise channel as:

\begin{equation}
    \Lambda_N(\rho) = (1- \delta) (p'_N \rho + (1-p'_N) \frac{I}{d}) + \delta \rho'
\end{equation}

for some parameter $p'_N$ (note that this is not equal to the depolarising parameter $p_N$ of the equivalent depolarising channel), where $\rho'$ is a state that potentially depends on $\rho$.

The left hand side of Eq.~\ref{eq: proof} is upper bounded by

\begin{eqnarray}
    ||\Lambda_N - \Lambda_{N_d}||_{\diamond} + ||\Lambda_N - \Lambda_{N_d}||_{\diamond} + ||\Lambda - \Lambda_{d}||_{\diamond} \\
    \leq 2\delta +  ||\Lambda - \Lambda_{d}||_{\diamond}
\end{eqnarray}

which is upper bounded by  $ 2\delta + E'$ 

Now all the quantities in Eq.~\ref{eq: proof} are known from experiment or are bounded. If we set $p^{est}_N = \sqrt{\frac{p_{\overline{C}}}{p}}$ we have that:

\begin{equation}
    |  p^2_N - p^{est \text{ } 2}_N  | \leq \frac{d}{d-1} \frac{E' + 2 \delta}{p} 
\end{equation}

Or following the same steps as in the Sec.~\ref{sec: noiseanalysis} we have:

\begin{eqnarray}
    | r_N - r_N^{est} | \leq  \sqrt{\frac{d-1}{d} \frac{E' + 2 \delta}{p}} 
\end{eqnarray}

If noise in the non-Clifford gate is Pauli noise, we know that $||\Lambda_N - \Lambda_{N_d}||_{\diamond} \leq \frac{2(d^2 -1)(1-p)}{d^2}$

so the same derivation as above gives us:

\begin{eqnarray}
    | r_N - r_N^{est} | \leq  \sqrt{\frac{d-1}{d} \frac{E''}{p}}
\end{eqnarray}

where  $E''=  \frac{6(d^2 -1)(1-p)}{d^2} + 4 \sqrt{1-p} \sqrt{d^2-1}$

If we know the bound for the diamond norm distance from a particular noise model to the depolarising channel then we can apply our treatment.

\end{document}